\documentclass{jetpl}
\usepackage{amsmath,amsfonts,graphics,epsfig,amssymb}
\usepackage[english]{babel}
\usepackage{hyperref}
\usepackage{xcolor} %% for color comments
\usepackage{ulem} %% for strike-out
\definecolor{myblue}{rgb}{0.05,0.1,0.5}

\twocolumn
\lat

\title{
Upper limits on the isotropic diffuse flux of cosmic PeV photons from \textit{Carpet-2} observations
}

\rtitle{Diffuse cosmic PeV photons}

\sodtitle{Upper limits on the isotropic diffuse flux of cosmic PeV photons from \textit{Carpet-2} observations}

\author{~\\[-7mm]
D.\,D.\,Dzhappuev$^{a}$,
Yu.\,Z.\,Afashokov$^{a}$,
I.\,M.\,Dzaparova$^{a,b}$,
T.\,A.\,Dzhatdoev$^{a,c}$,
E.\,A.\,Gorbacheva$^{a}$,
I.\,S.\,Karpikov$^{a}$,
M.\,M.\,Khadzhiev$^{a}$,
N.\,F.\,Klimenko$^{a}$,
A.\,U.\,Kudzhaev$^{a}$,
A.\,N.\,Kurenya$^{a}$,
A.\,S.\,Lidvansky$^{a}$,
O.\,I.\,Mikhailova$^{a}$,
V.\,B.\,Petkov$^{a,b}$,
E.\,I.\,Podlesnyi$^{c,a}$,
N.\,A.\,Pozdnukhov$^{c,a}$,
V.\,S.\,Romanenko$^{a}$,
G.\,I.\,Rubtsov$^{a}$,
S.\,V.\,Troitsky$^{a,c}$\thanks{Corresponding author; e-mail:
st@ms2.inr.ac.ru},
I.\,B.\,Unatlokov$^{a}$,
I.\,A.\,Vaiman$^{a}$,
A.\,F.\,Yanin$^{a}$,
K.\,V.\,Zhuravleva$^{a}$\\
(Carpet--2 Group)\\
~}
\rauthor{Carpet--2 Group}

\sodauthor{D.\,D.\,Dzhappuev et al.\ (Carpet--2 Group)}

\address{
$^{a}$Institute for Nuclear Research of the Russian Academy of
Sciences,\\
60th October Anniversary prospect 7A, 117312 Moscow, Russia\\
$^{b}$Institute of Astronomy, Russian Academy of Sciences, Moscow, 119017
Russia\\
$^{c}$M.V. Lomonosov Moscow State University, 1-2 Leninskie Gory,  Moscow 119991, Russia}

\dates{December 2, 2022}{*}

\abstract{Isotropic diffuse gamma-ray flux in the PeV energy band is an important tool for multimessenger tests of models of the origin of high-energy astrophysical neutrinos and for new-physics searches. So far, this flux has not yet been observed. Carpet-2 is an air-shower experiment capable of detecting astrophysical gamma rays with energies above 0.1~PeV. Here we report the upper limits on the isotropic gamma-ray flux from Carpet-2 data obtained in 1999--2011 and 2018--2022. These results, obtained with the new statistical method based on the shape of the muon-number distribution, summarize Carpet-2 observations as the upgraded installation, Carpet-3, starts its operation.
}
\begin{document}
\maketitle

\noindent\textbf{1. Introduction.} 
Searches for PeV gamma rays flourished since the discovery of astrophysical neutrinos of similar energies was announced \cite{IceCube-2013}. Indeed, in the conventional scenario, neutrinos are born from charged $\pi$-meson decays, while neutral pions produce gamma rays. Pair production on the cosmic microwave background \cite{Nikishov} limits the mean free path of PeV photons to tens of kiloparsecs, that is the size of our Galaxy. Search for these photons, therefore, is important to distinguish between Galactic and extragalactic origin of high-energy neutrinos \cite{AhlersMurase-gamma-Gal,Winter-gamma-Gal,KalashevST-gamma-Gal}. In particular, several Galactic models do not produce strong anisotropy of neutrino arrival directions because the neutrinos are born either in the neighbourhood of the Solar system \cite{NeronovBubble2018,LocalBubble2020} or in the spherically symmetric Galactic halo \cite{Taylor:halo-neutrino,KT2016,Gabici:2021-crhalo,KMT2022}. These models can be tested with the search for (sub)-PeV isotropic diffuse gamma rays. In addition, this kind of energetic gamma-ray background may appear in new-physics models where the pair production is suppressed, e.g.\ because of the axion-photon oscillations or Lorentz-invariance violation, see e.g.\ \cite{ST-mini-rev,LIVrev} and references therein. In models of Lorentz violation, however, photons may evade registration due to the suppression of the interaction cross section in the atmosphere \cite{noLIVeas1,noLIVeas2}. Recent observations reported anisotropic diffuse flux of sub-PeV gamma rays from the Galactic plane \cite{Tibet-GalDiffuse}. Here, we obtain upper limits on the isotropic photon flux from Carpet-2 data. We present a new method to search for primary photons, which does not rely on the simulations of cosmic-ray background, as well as results of its application.

\vskip 2mm
\noindent\textbf{2. Data.}
\vskip 0.5mm
\noindent{\sl 2.1. Experiment.} Carpet-2 is a surface air-shower array located at the Baksan Neutrino Observatory of the Institute for Nuclear Research of the Russian Academy of Sciences (Neutrino village, North Caucasus; geographical coordinates 43.273$^\circ$N, 42.685$^\circ$E). It contains a 200~m$^2$ continuous central detector used to estimate the effective number $N_e$ of relativistic particles in the shower core, five external stations used to determine the shower arrival direction, and a 175~m$^2$ muon detector which records the number $n_\mu$ of muons (muon energy threshold of 1~GeV for vertical particles). Details of the installation, data acquisition and processing are described elsewhere \cite{Carpet2-description2007,Carpet2-Szabelski2009,Carpet2-arXiv2015,Carpet2019ICRC...36..808T,Carpet2019TeVPA-point}. Among recent results of the Carpet-2 experiment are the observation of a flare of gamma radiation above 300~TeV associated with a 150-TeV neutrino from a Galactic source \cite{Carpet-ApJL2021}, the detection of a photon-like event with estimated primary energy of 251~TeV associated with GRB~221009A \cite{CarpetATel-GRB} and possibly indicating new particle physics, see e.g.\ \cite{GRB-ALP1,GRB-ALP2}, as well as constraints on sub-PeV photons from directions of IceCube and HAWC alerts \cite{Carpet2-IceCubeHAWC}.
\vskip 0.5mm
\noindent{\sl 2.2. Two data sets.}
Since the beginning of operation of the muon detector in 1999, Carpet-2 data were collected in two periods. Before 2011, the installation worked with the main goal of studying cosmic-ray showers, and the trigger condition $n_\mu>1$ was imposed. Therefore, during this long-exposure period, the efficiency of the installation to muon-poor photon-induced showers was low. The installation resumed operation in 2018 with the aim to search for gamma rays, therefore the trigger condition was changed to include all events independently of $n_\mu$. For the purpose of this work, we use two data sets:
\begin{itemize}
    \item 
    \textit{Maximal-exposure} data set which includes the combination of 1999-2011 events and 2018-2022 events with the cut $n_\mu>1$ imposed in addition to the standard quality cuts \cite{Carpet2019JETPLneutrino-old}. The set consists of 213691 events.
    \item 
    \textit{Photon-friendly} data set which includes the events recorded in 2018-2022 without the $n_\mu$ cut. The set consists of 104527 events.
\end{itemize}
The last day of data included is November 6, 2022.

\vskip 0.5mm
\noindent{\sl 2.3. Monte-Carlo simulations.}
The present work uses Monte-Carlo (MC) simulations of photon-induced air showers and the Carpet-2 detector response described in Ref.~\cite{Carpet2019JETPLneutrino-old}. Simulations assume isotropic gamma-ray flux with primary energies above 30~TeV. The response of the installation to each air shower is simulated with a dedicated code, and the resulted artificial events pass exactly the same selection and reconstruction procedures as the real data. 

\vskip 1mm
\noindent\textbf{3. Constraining the photon flux.} 
It is customary to use the muon content of air showers to distinguish rare events caused by primary photons from the bulk of hadron-induced showers. This approach is based on the fact that the probability of a photonuclear or electronuclear interaction is suppressed with respect to a purely electromagnetic reaction for energetic photons and electrons. As a result, showers caused by primary gamma rays are poor in muons, and a low value of the ratio $n_\mu/N_e$ becomes a useful tracer of photon-induced events. Previous studies used particular criteria to select ``photon candidates'' with low $n_\mu/N_e$, and the photon fluxes were estimated from the number of these candidate events with the account of both the amount of gamma-ray showers which were not selected by the criteria and the amount of hadronic showers which passed the criteria to become photon candidates. The optimal $n_\mu/N_e$ cut and the selection efficiency for both photons and hadrons were determined from MC simulations. However, this approach has a serious source of uncertainties, because the predicted muon content of hadronic air showers depends strongly on the assumed primary composition, spectra and hadronic-interaction models. Since we are interested in the very tail of $n_\mu/N_e$ distribution for hadronic showers to estimate the background of muon-poor events not related to primary photons, these uncertainties affect strongly the robustness of the results obtained in that approach. 

\begin{figure}
    \centering
   \includegraphics[width=0.9\linewidth]{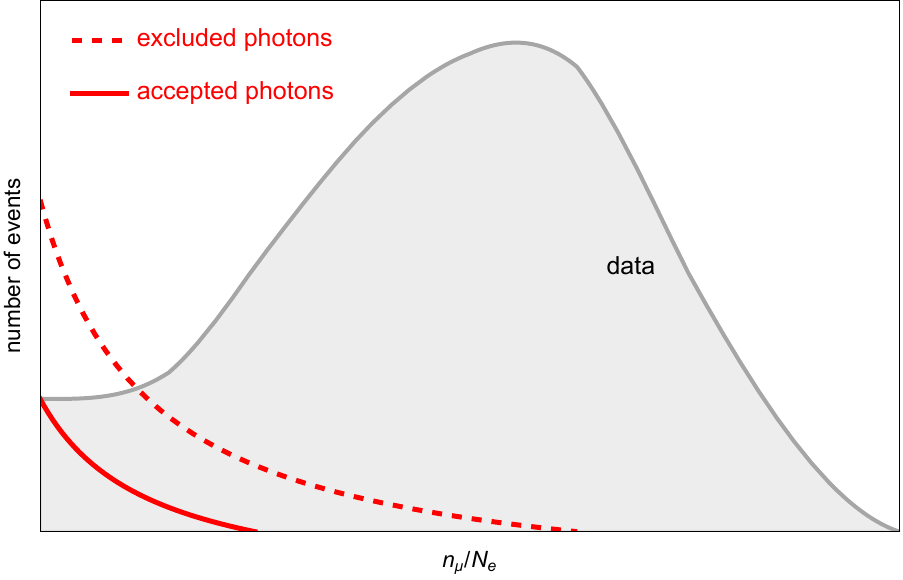}
    \caption{\label{fig:sketch} \sl
\textbf{Figure~\ref{fig:sketch}}
Sketch (arbitrary units; not based on real data or simulations) of the idea of the statistical method to constrain the photon flux from the shape of $n_\mu/N_e$ distribution.
}
\end{figure}

To avoid these uncertainties, we use here a new statistical method which does not rely on MC simulations of the hadronic background, nor on particular criteria for photon candidates. Instead, it exploits the shapes of the distributions of electromagnetic and hadronic showers in $n_\mu/N_e$, which are very different. The idea of the approach is illustrated by a sketch in Fig.~\ref{fig:sketch}. 
While the tail of the $n_\mu/N_e$ distribution in data extends all the way to $0$, its peak is well separated from 0 because the dominant part of the detected events, if not all of them, are produced by hadrons (shaded region in Fig.~\ref{fig:sketch}). Contrary, simulated $n_\mu/N_e$ distributions for primary photons peak at 0 since in the energy range of interest, most of the gamma-ray events are muonless. For sufficiently small number of photon-induced events in the sample, their contribution does not contradict the observed distribution (full red line in Fig.~\ref{fig:sketch}), while larger photon contributions (like the one corresponding to the dashed line) are excluded.

\begin{figure}
    \centering
   \includegraphics[width=0.9\linewidth]{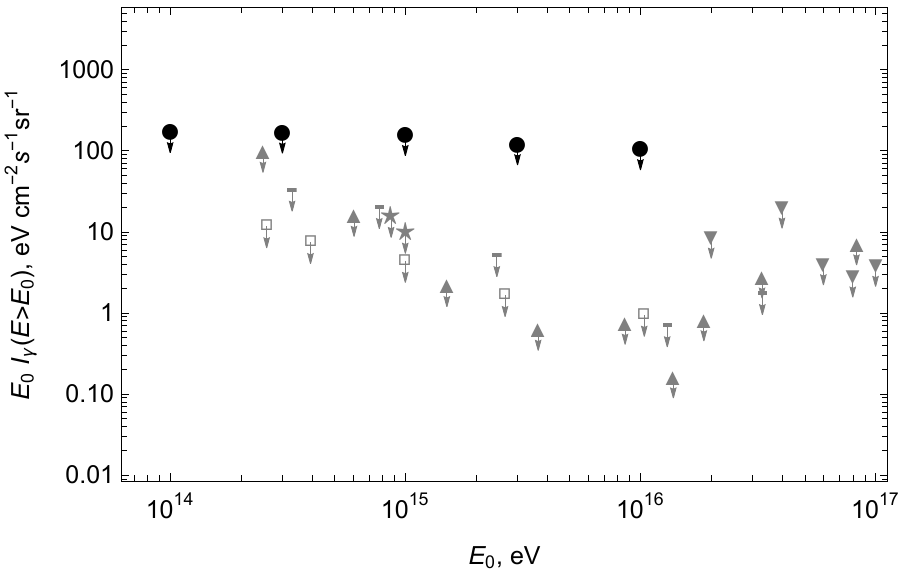}
    \caption{\label{fig:integral} \sl
\textbf{Figure~\ref{fig:integral}.}
90\% CL upper limits on the integral isotropic diffuse flux
of high-energy photons. Black circles: Carpet-2, this work ( Table~\ref{tab:integral}; strongest limits of the two data sets). Gray symbols ---
limits from other experiments 
(empty boxes --
KASCADE~\cite{KASCADE-2003-gamma}, upward triangles -- KASCADE and
KASCADE-Grande~\cite{KASCADE-2017-gamma}, downward triangles --
EAS-MSU~\cite{EAS-MSU-gamma}, horizontal dashes --
CASA-MIA~\cite{CASA-MIA-gamma}, asterisks -- EAS-TOP~\cite{EAS-TOP-gamma}).}
\end{figure}

\begin{figure}
    \centering
   \includegraphics[width=0.9\linewidth]{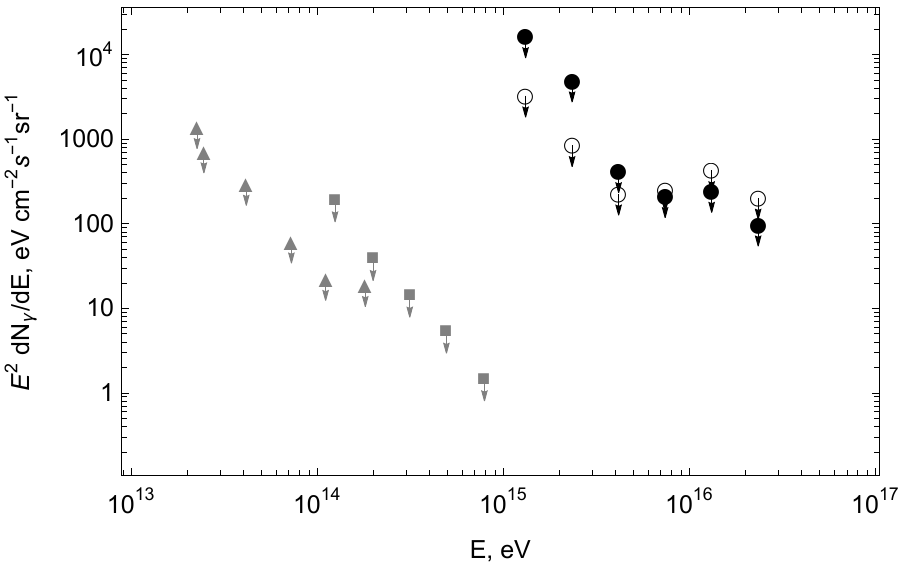}
    \caption{\label{fig:differential} \sl
\textbf{Figure~\ref{fig:differential}}
90\% CL upper limits on the differential isotropic diffuse flux
of high-energy photons. Black symbols: Carpet-2, this work (Table~\ref{tab:differential}; full circles -- maximal-exposure data set, empty circles -- photon-friendly data set). Gray symbols ---
limits from other experiments 
(triangles -- HAWC \cite{HAWC-diffuse-limits},
squares -- analysis of Tibet-AS$\gamma$ results by other authors
\cite{Neronov:2021ezg}).}
\end{figure}

In practice, these distributions are not smooth since they are produced by a finite number of events, and fluctuations are important, especially when testing low fractions of photons. The practical statistical implementation of our approach to the search of photons based on the shape of the $n_\mu/N_e$ distribution consists of the following five steps.
\begin{enumerate}
\item
Fix the desired range of photon energies $E$ and determine the corresponding range of reconstructed $N_e$ using MC for primary photons, assuming the $E^{-2}$ spectrum.
\item
Select events corresponding to this $N_e$ range and passing all quality cuts both from the real data and from photon MC. Determine the efficiency $\epsilon$ as the ratio of the number of photon MC events passed the quality cuts and the $N_e$ selection to the number of photon MC events with thrown energies in the fixed $E$ range.
\item
Obtain the binned distribution of the selected data in $\log \left(n_\mu/N_e \right)$. The bin size is fixed to 0.2 as this corresponds to the precision of determination of this observable, determined from MC simulations\footnote{We have tested explicitly that reasonable variations of the bin size do not affect the result.}. A separate bin is reserved for events with $n_\mu=0$.
\item
Introduce the variable $N_\gamma$, the Poisson mean for the number of photon-induced events in the data set. For different values of $N_\gamma$, generate many ($10^4$) realisations of simulated photons, selected randomly from the MC pool determined at step~2. The number of events is determined every time from $N_\gamma$ with the help of the Poisson distribution. For each realisation, obtain the $n_\mu/N_e$ distribution similar to that obtained for the real data at step~3. If this simulated distribution does not exceed the real-data distribution in all bins (cf.\ the thick ``accepted'' curve in Fig.~\ref{fig:sketch}), count this realisation as ``accepted''. 
\item
The fraction of ``accepted'' realisations decreases with $N_\gamma$ increasing. Because of fluctuations, the presence of ``not accepted'' cases does not mean the corresponding value of $N_\gamma$ is excluded. Find the upper limits $N_{\gamma,90}$ ($N_{\gamma,95}$), for which 90\% (95\%) of realizations are accepted. Obtain the 90\% CL (95\% CL) upper limits on the allowed gamma-ray flux from $N_{\gamma,90}$ ($N_{\gamma,95}$), the efficiency $\epsilon$ determined at step 2, and the exposure of the installation.
\end{enumerate}
\vskip 1mm
\noindent\textbf{4. Results.}
It is customary to report upper limits on the \textit{integral} diffuse gamma-ray flux at energies above PeV but limits or measurements of the \textit{differential} flux below PeV. Since Carpet-2 tests the PeV range, we report our results in both ways. The $E^{-2}$ spectrum of primary photons is assumed for the entire energy range for integral-flux limits and for each energy bin for differential-flux limits. The results for the differential and integral fluxes are presented in Tables~\ref{tab:integral}, \ref{tab:differential},  and compared to those published by other groups in Figures~\ref{fig:integral}, \ref{fig:differential}, respectively.
\begin{table*}
\begin{center}
\begin{tabular}{ccccc}
\hline
\hline
 $E_{\rm min}$,  & 
\multicolumn{4}{c}{$EI(E>E_{\rm min}$) upper limits, eV/cm$^2$/s/sr}
\\
TeV & \multicolumn{2}{c}{maximal-exposure set} & \multicolumn{2}{c}{photon-friendly set}\\
&95\% CL & 90\% CL & 95\% CL & 90\% CL\\
\hline
  100    & 199 & 174 & 415 & 340\\
  300    & 196 & 172 & 433 & 359\\
 1000    & 183 & 160 & 385 & 315\\
 3000    & 164 & 142 & 150 & 123\\
10000    & 129 & 107 & 227 & 183\\
\hline
\hline
\end{tabular}
\end{center}
\caption{\label{tab:integral}
Table~\ref{tab:integral}:
Upper limits on the integral isotropic diffuse flux
of high-energy photons obtained in the present work. $E^{-2}$ spectrum is assumed for all energies. 
}
\end{table*}
\begin{table*}
\begin{center}
\begin{tabular}{ccccccc}
\hline
\hline
\multicolumn{2}{c}{$\log_{10}(E/\mbox{TeV})$}    & $\langle E\rangle$,  & 
\multicolumn{4}{c}{$E^2 dN/dE$ upper limits, eV/cm$^2$/s/sr}
\\
min      & max &TeV & \multicolumn{2}{c}{maximal-exposure set} & \multicolumn{2}{c}{photon-friendly set}\\
&&&95\% CL & 90\% CL & 95\% CL & 90\% CL\\
\hline
3.0  & 3.25 &1315    & 17100 & 16400 & 3360 & 3260\\
3.25 & 3.5  &2339    &  5080 &  4910 &  926 &  845\\
3.5  & 3.75 &4159    &   482 &   416 &  280 &  226\\
3.75 & 4.0  &7396    &   260 &   211 &  303 &  247\\
4.0  & 4.25 &13153   &   309 &   243 &  528 &  427\\
4.25 & 4.5  &23390   &   127 &  97.5 &  264 &  202\\
\hline
\hline
\end{tabular}
\end{center}
\caption{\label{tab:differential}
Table~\ref{tab:differential}:
Upper limits on the differential isotropic diffuse flux
of high-energy photons obtained in the present work. $E^{-2}$ spectrum is assumed in each energy bin. $\langle E\rangle$ is the mean energy for the bin under this assumption.
}
\end{table*}

As one could expect, differential-flux limits from the photon-friendly data set are stronger at lower energies while those from the maximal-exposure data set win at high energies. These high-energy bins dominate also the integral-flux limits, so in most cases these limits are stronger if derived from the maximal-exposure data set.

Differential-flux limits on the isotropic diffuse gamma-ray flux above 1~PeV have not been published previously by any experiment. For the integral flux, the limit at $E>100$~TeV is also new. The GRAPES-3 experiment has reported~\cite{GRAPES3-gamma} on the gamma-ray fraction limits in this energy range; however, conversion of the fraction limits to flux limits is ambiguous. 

\vskip 1mm
\noindent\textbf{5. Summary.}
We present a novel method to search for primary photons in an air-shower experiment, Carpet-2, using the shape of the distribution of events in a primary-discriminating observable. This approach avoids uncertainties related to modelling of the background of cosmic-ray induced showers at the tails of the distribution. The approach can easily be generalized and applied to the data of other installations capable of detecting muons in air showers, e.g.\ Yakutsk \cite{Yakutsk}, NEVOD \cite{NEVOD} etc. We apply this method to the full Carpet-2 data set and report the obtained limits on the diffuse gamma-ray background. These limits may be used within the multimessenger approach to search for high-energy astrophysical neutrino origin or for manifestations of new physics. 

The installation starts to operate with the extended muon detector of 410~m$^2$ in 2022 and will soon be upgraded to Carpet-3, covering a much larger surface area compared to the present data. A tenfold increase of exposure to high-energy gamma-rays is expected. With future large-scale installations, like LHAASO \cite{LHAASO} and SWGO \cite{SWGO}, the diffuse isotropic flux of PeV gamma rays might be eventually discovered. 

\vskip 1mm
\noindent\textbf{Acknowledgements.} This work was supported by the RF Ministry of science and higher education under the contract 075-15-2020-778.

\bibliographystyle{nature1}
\bibliography{diffuse}
\end{document}